\documentclass[pra,12pt,english]{article}
\usepackage{graphicx}
\begin{document}
\title{The Elitzur-Vaidman  Interaction-Free Measurements}

\maketitle

\vspace {.1cm}

The interaction-free measurements proposed by Elitzur and Vaidman
\cite{EV93} (EV IFM) is a quantum mechanical method to find an
object that interacts with other systems solely via its explosion
without exploding it. In this method, an object can be found without
``touching it'', i.e. without any particle being at its vicinity.

The basic idea of the method is as follows. A quantum test particle
is being split into a superposition of two separated states. One of
these state is being split again into a superposition of two output
states while the other is being split into a (different)
superposition of the same output states. The phases of the various
parts are tuned in such a way that there is a destructive
interference at one of the outputs. At this output there is a
detector. This is the EV device ready for action.

The simplest EV device is the Mach-Zehnder interferometer, Fig. 1.
To use it, the device should be placed in such a way that only one
of the intermediate states interacts with the object.  If the object
is present, the destructive interference is spoiled and the detector
might click announcing that the object is present. In this case, no
explosion has occurred, since the particle can be found only in one
place. The particle can also be ``found'' by the object, so in half
of the cases the object explodes. The probability of finding the
object on the first run is just one quarter, so the efficiency of
the method is low, but given that the detector clicks, the object is
present with certainty.

\begin{figure}
  \includegraphics[width=22cm]{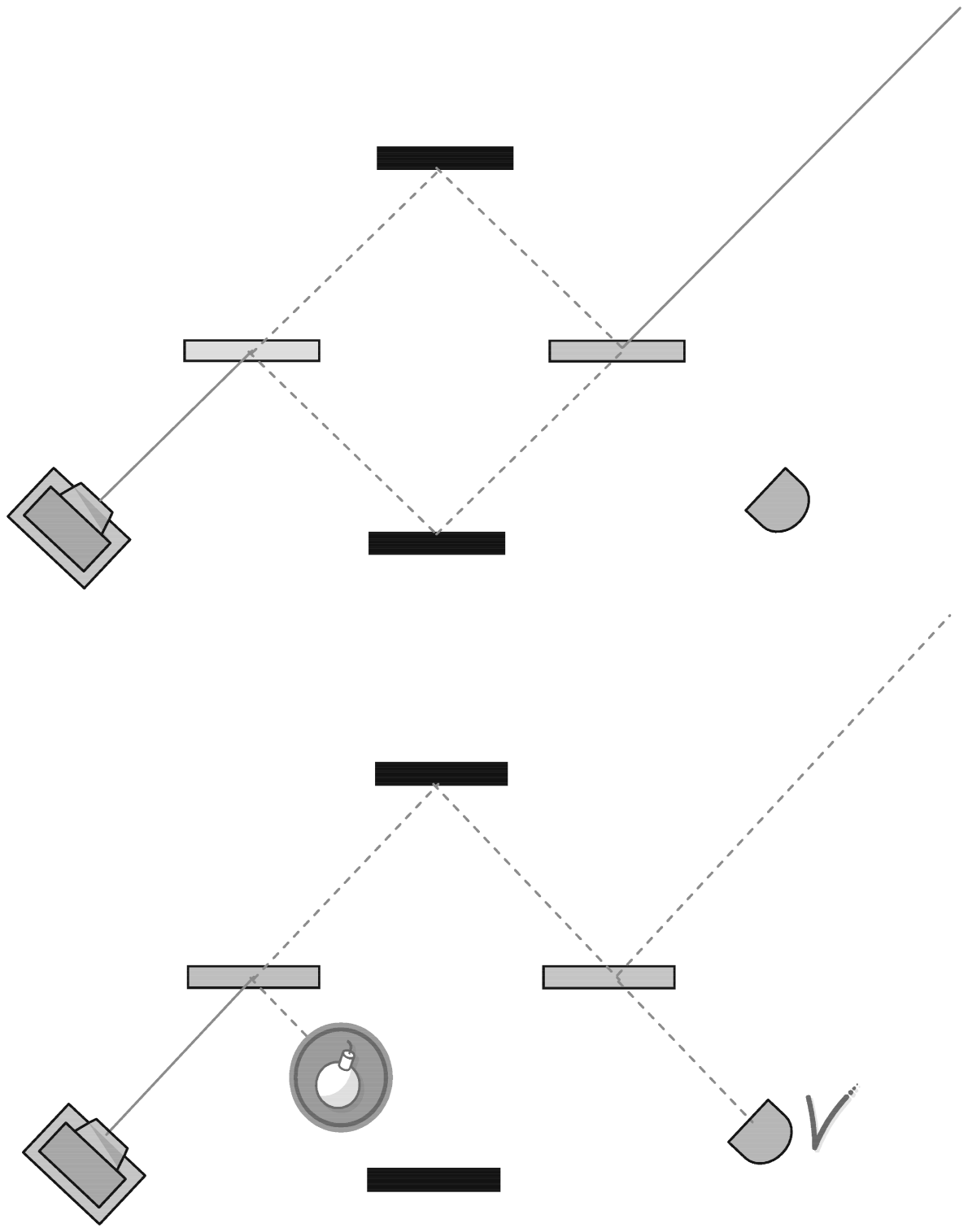}\\
  \caption{ {\bf The Elitzur-Vaidman scheme}~
\hfill\break
 (a) When  the interferometer is empty and
  properly tuned, photons do not reach the detector.
 \hfill\break (b) If the exploding object is present, the detector has the
  probability 25\% to detect the photon  sent through the
  interferometer, and in this case we know that the object is inside the
  interferometer without exploding it.}\label{yy}
\end{figure}

The EV method was improved using the Zeno effect \cite{Kwi95} and
the probability of the explosion could be made arbitrary small.
This, however, requires more time: the quantum test particle has to
traverse the interaction region many times. Conceptually, the
simplest implementation of this improvement is a device consisting
of two identical cavities $A$ and $B$ connected by a highly
reflective wall, see Fig 2. If we place a photon in one cavity, the
evolution brings it to another cavity after $N$ bounces in one
cavity . At this moment, a detector tests for the presence of the
photon in cavity $A$. This is the device which is ready for action.
We place it in such a way that the interaction region of possible
explosive object is cavity $B$. The detector will click with
probability close to 1. (The probability for the failure, which is
an explosion of the object, is of the order of $1\over N$). It will
not click for sure if the object is absent.

Setups similar to the EV device were considered before by Renninger
\cite{Ren53} and Dicke \cite{Dick}. However, they did not realized
the effect because in their analysis the object and the test
particle were reversed: they pointed out the peculiar property that
the EV test particle changes its state while the EV explosive object
(their measuring device) has not changed at all, it was a {\it
negative result experiment}.

The EV method can find in an interaction-free manner not only
exploding objects, but any opaque object. This experiment, however,
is somewhat more difficult to implement. For finding an explosive
device we could use, instead of a single particle source, a weak
laser beam. If the click happens before the explosion, we know that
the object is there. For an opaque object, we need a single particle
source: if we get a click sending only one photon, we know that
there is a opaque object somewhere inside the interferometer and
that it did not absorb any photon.

\begin{figure}
  \includegraphics[width=12cm]{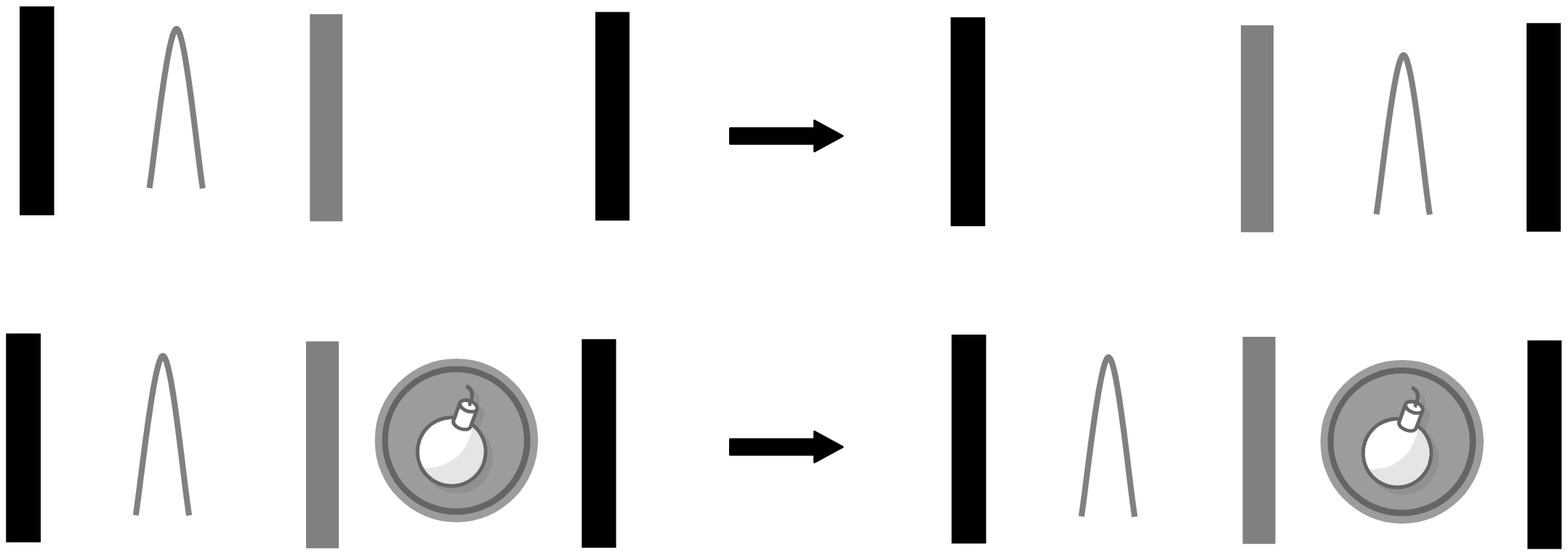}\\
  \caption{ {\bf The Kwiat et al. scheme.} \hfill\break~(a)If the
cavities are empty, the photon after $N$ bouncing moves completely
from the left cavity to the right cavity. \hfill\break (b) If the
object is present in the second cavity, after the same $N$ bounces
will remain in the first cavity with probability close to 1 for
large $N$.}\label{yyy}
\end{figure}

One of the most paradoxical features of the EV IFM is that the test
particle in some sense never passes in the vicinity of the
interaction region. How can we get information about a region when
nothing passed through it and nothing came out of it? Indeed, when
we hear the click announcing the presence of the object, there is no
record of any kind in our world  showing that the test particle was
near the object.

A way to resolve this paradox is to note that of our intuition
regarding causality in our world is based on physical laws. These
laws, however, describe our Universe which includes many worlds,
including the one in which the test particle visited the interaction
region (and there was an explosion). In this picture it is easy to
understand why there is no interaction free method for finding out
that the interaction region is empty. Since there is no parallel
world in which an explosion occur, we cannot verify that the region
is empty without passing through it.

Let us consider now what happens when the EV IFM device is used for
finding a quantum object. If the wave function of the quantum object
spreads over space such that only part of it overlaps with the
interaction region, the  successful EV IFM  localizes the object to
the interaction region without changing its internal state (without
exploding it). The momentum of the object is changed in this
procedure. In this respect it is no different from any other
nondemolition measurement of the projection on the interaction
region.  The name ``energy exchange free measurement'' frequently
associated with the EV proposal, thus does not reflect the unique
features of the EV IFM \cite{mean}.

Energy exchange is relevant for the Penrose modification of the EV
IFM \cite{Pen}, in which the goal is different: We are to
distinguish between objects which explode whenever their trigger is
touched and duds where the trigger is locked to the object which do
not explode. The dud serves as a mirror in the Mach-Zehnder
interferometer which produces a destructive interference in its
detector. A good exploding device cannot server as a mirror and thus
the detector might click announcing that the object is not a dud.
Penrose's explanation of the core of the IFM is {\it counterfactual}
\cite{CFentry}: the object caused the detector to click because it
could have explode, although it did not. This is the origin of the
name {\it counterfactual computation} \cite{MiJo,Ho} for a quantum
computer which yields the outcome without ``running'' the algorithm.
Note, however, that as we cannot establish the {\it absence} of an
object in an interaction-free manner, we cannot have a
counterfactual computation for all possible outcomes \cite{CFCV}.

In Penrose's IFM, when the detector clicks, we can claim, as before,
that the quantum test particle was not at the vicinity of the
exploding object. However, when the EV IFM device is used for
finding a quantum object,  the click of the detector  does not
ensure that the quantum test particle was not present in the
interaction region. It might  that the whole quantum wave of the
test particle passes the interaction region. This happens when the
observed quantum object is the ``test particle'' of the EV IFM
measuring the presence of the original test particle. This setup is
known as the Hardy's paradox \cite{Hardy}. This consideration shows
that the claim that the EV IFM localizes quantum objects to the
interaction region is strictly speaking incorrect. But limitation is
minor: anyone observing the location of the object (and not a
superposition of localized states) after the EV IFM announcement
about its location, will find that EV IFM method is not mistaken.

 There have been numerous experiments performing the EV IFM. The
original EV scheme was  first implemented in laboratory by Kwiat et
al. \cite{Kwi95}. Later, Kwiat et al. also performed an experiment
of their improved scheme which combines the EV setup with the Zeno
Effect \cite{Kwi99} reaching efficiency of about  70\%. Technical
problems make further improvement  difficult. It is not easy to tune
the optical cavities and it is very difficult to put the photon into
the first cavity at a particular moment for starting the process.

When the goal is a practical application of the EV IFM, the best
approach is the Paul and Pavi\v ci\'c setup \cite{PaPa} which is,
essentially a Fabry Perot interferometer, Fig. 3.  There is only one
cavity build with almost 100\% reflecting mirrors, which is tuned to
be transparent when empty.If,  however, there is an object inside
the cavity, it becomes almost 100\% reflective mirror which allows
finding the object without exploding it. The method has a conceptual
drawback that in principle the photon can be reflected even if the
cavity is empty, thus, detecting reflected photon cannot ensure
presence of the object with 100\% certainty. But this drawback has
no meaning for actual experiment because noise in an ideal setup is
usually larger. This method was first implemented in a laboratory by
Tsegaye {\it et al.} \cite{Tse98} and recent experiment reached the
efficiency of 88\% \cite{NaIn}. The method has a potential to
improve controlled-not gate for quantum information processing
\cite{Pa}.

\begin{figure}
  \includegraphics[width=12cm]{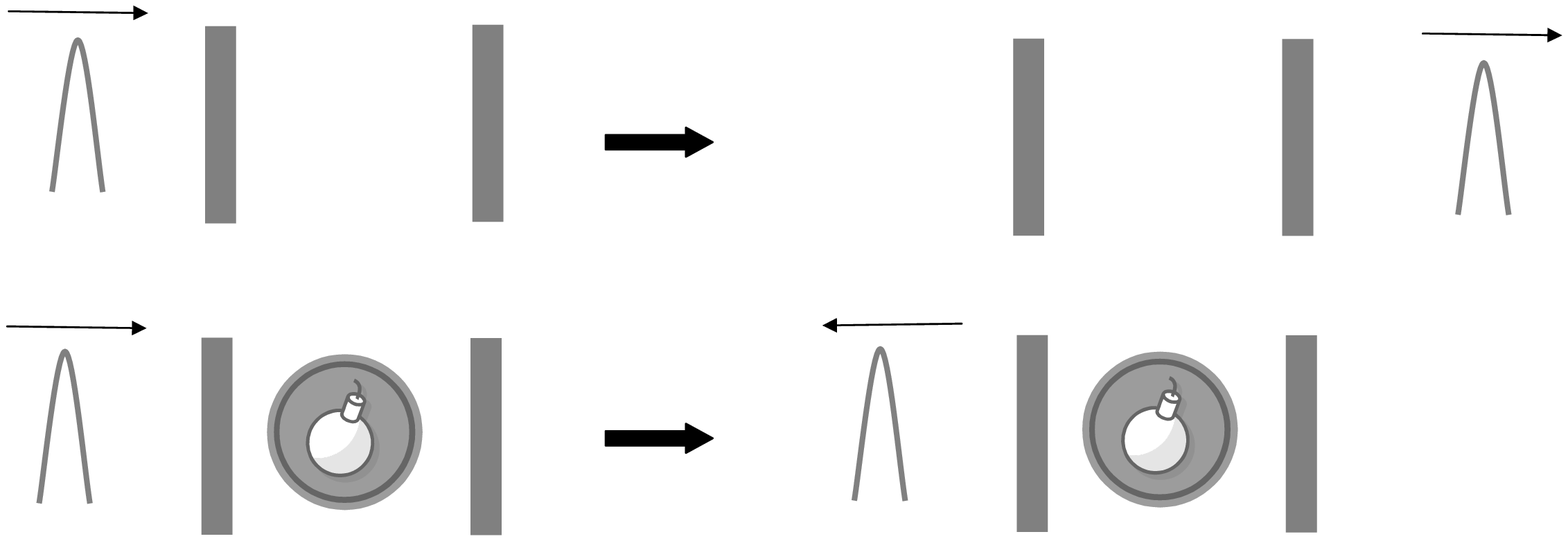}\\
  \caption{ {\bf The Paul and Pavi\v ci\'c
scheme.}~\hfill\break~(a) If the cavity is empty, the photon passes
through it with very high probability. \hfill\break (b) If the
object is present in the cavity, the photon is reflected with very
high probability.}\label{yyy}
\end{figure}

Applying the EV device for imaging semitransparent objects
\cite{Jang,MiMa,Azum} hardly pass the strict definition of the IFM
in the sense that the photons do not pass in the vicinity of the
object, but they achieve a very important practical goal, since we
``see'' the object significantly reducing  the irradiation of the
object: this can allow measurements on fragile objects.

The EV IFM is one of the ``quantum paradoxes''. It is a task which
cannot be performed in the realm of classical physics, but can be
done in the framework of quantum theory. Progress in  experimental
demonstrations of the method shows that it has a potential for
practical applications.

This work has been supported in part by the European Commission
under the Integrated Project Qubit Applications (QAP) funded by the
IST directorate as Contract Number 015848 and by grant 990/06 of the
Israel Science Foundation.

\vskip .5cm Lev Vaidman\hfill\break The Raymond and Beverly Sackler
School of Physics and Astronomy\hfill\break Tel-Aviv University,
Tel-Aviv 69978, Israel


\end{document}